\documentclass[aps,12pt,notitlepage]{revtex4-1}
\usepackage{bm}
\usepackage{color} 
\usepackage{graphicx}
\usepackage{epsfig} 
\usepackage{amsmath}
\usepackage{amssymb}
\usepackage{float}
\begin{document}

\title{ Thermal generation of shift electric current}

\author{G.\,V.\,Budkin} 
\author{S.\,A.\,Tarasenko}

\affiliation{Ioffe Institute, 194021 St.~Petersburg, Russia} 

\begin{abstract}
It is shown that the dissipation of energy in an electron gas confined in a quantum well made of non-centrosymmetric crystal leads to a direct electric current. The current originates from the real-space shift of the wavepackets of Bloch electrons at the electron scattering by phonons, which tends to restore thermal equilibrium between the electron and phonon subsystems. We develop a microscopic theory of such a phonogalvanic effect for narrow band gap zinc-blende quantum wells.
\end{abstract}   
    
\maketitle
 
\section{Introduction}   

The currents of charge carriers, electrons or holes, in solid-state systems are commonly generated by gradients of electric or chemical potentials, temperature, etc. 
In structures of low spatial symmetry, direct currents can 
also emerge when the system is driven out of thermal equilibrium by an undirected force with zero average driving~\cite{feynmanbook,sturmanbook}.
Well known examples are ratchet and photogalvanic effects~\cite{reimann2002,hanggi2009,bang2018,hohberger2001,ivchenko2011,chepelianskii2008,tarasenko2011,drexler2013,budkin2016} and the spin-galvanic effect~\cite{ganichev2002,seibold2017,smirnov2017}, when non-equilibrium spin polarization drives an electric current in gyrotropic structures. Here, we study the effect of generating direct electric current from the kinetic energy of hot electrons. We show that the breaking of thermal equilibrium between the electron and phonon subsystems in semiconductor structure of sufficiently low symmetry is enough to drive an electric current. Microscopically, the current originates from the shift of Bloch electrons in real space at the emission or absorption of phonons, which tend to restore thermal equilibrium in the system.
While the spin-dependent shift of electrons at momentum scattering (side jump contribution) has been intensively studied, particularly, in the physics of the anomalous and spin Hall effects~\cite{luttinger1958,sinitsyn2007,nagaosa2010,oveshnikov2015,ado2017,xiao2019}, less is known about the charge shift which occurs at inelastic scattering~\cite{belinicher1982, golub2011}.
Here, we study this effect for narrow gap two-dimensional (2D) systems made of zinc-blend-type semiconductors, such as HgTe/CdHgTe quantum wells (QWs), which naturally lack the center of space inversion. These 2D structures, except for QWs grown along the high-symmetry axes [001] and [111], are polar in the plane and support the generation of electric current in the presence of energy transfer between the electron subsystem and the crystal lattice. In analogy to the (linear) photogalvanic effect~\cite{sturmanbook,baskin1978}, where the absorption of photons gives rise to a shift electric current, the effect we study can be named phonogalvanic.  

\section{Microscopic picture}

\begin{figure}[H]
\begin{center}
\includegraphics[width=0.5\paperwidth]{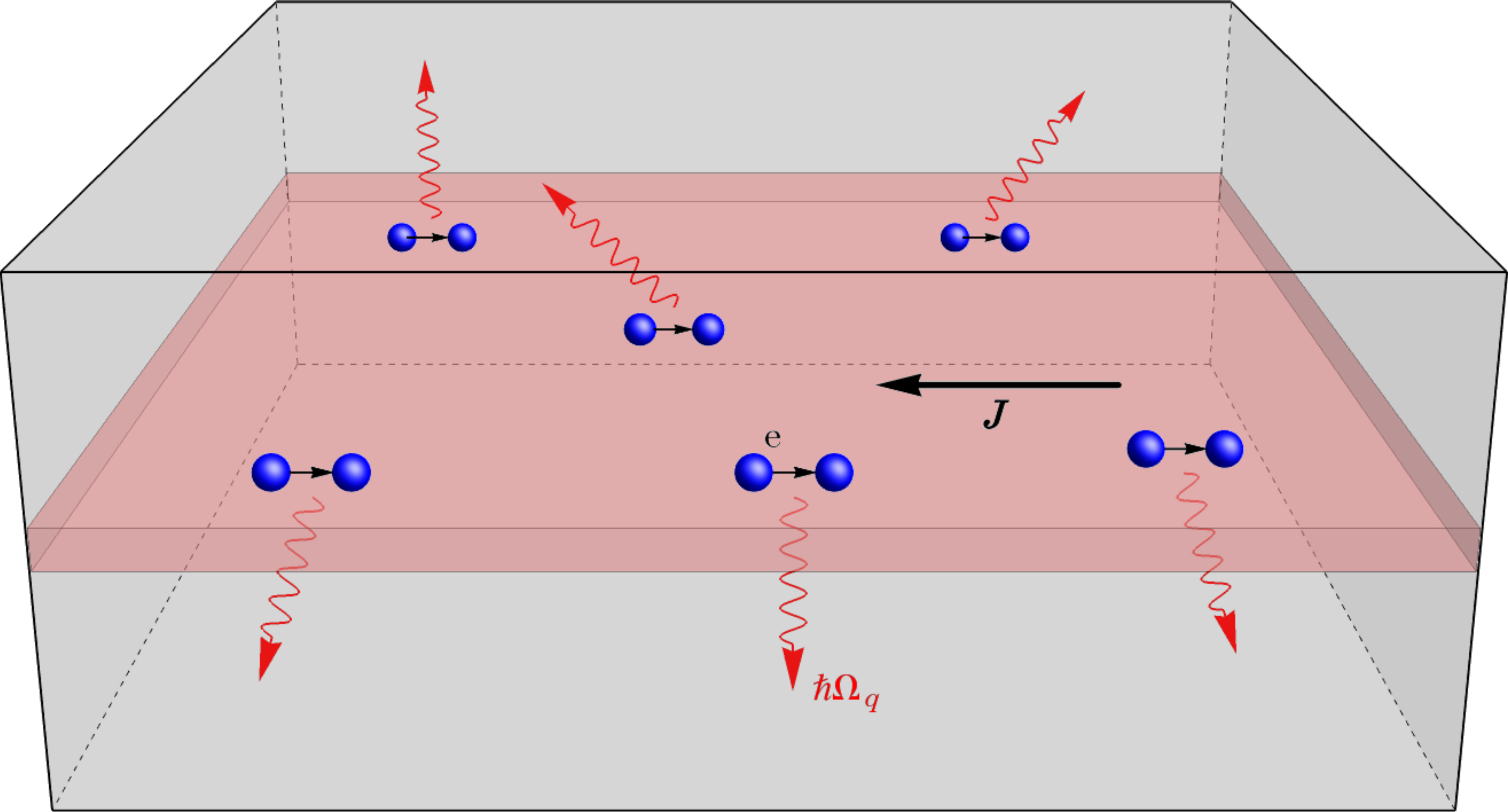}
\caption{Shift mechanism of phonogalvanic effect in quantum wells. Emission of phonons by hot electrons is accompanied by the electron shift in the QW plane, which leads to a net electric current.}
\label{figure_model}
\end{center}
\end{figure}

We consider a two-dimensional electron gas in a QW structure, see Fig.~\ref{figure_model}. The electron gas is initially heated. Frequent electron-electron collisions thermalize the electrons and establish the Fermi-Dirac distribution with the 
effective electron temperature $T_e$ that is above the lattice temperature $T$. The electron gas is being cooled down by emitting phonons which transfer the energy from the QW region. As will be calculated below, the processes of phonon emission/absorption in zinc-blend QWs are accompanied by the in-plane shift of the Bloch electrons involved in the quantum transitions.
The shifts have a preferable direction, except for the QWs grown along the high-symmetry axes [001] and [111], although the phonons can be emitted in all directions.
Therefore, the energy relaxation of electrons gives rise to a direct electric current $\bm j$. The current is generated until the electron temperate reaches the lattice temperature.

At thermodynamic equilibrium, the net current vanishes because the electron shifts at the phonon emission and absorption are opposite and compensate each other. If, however, the electron temperature is higher (or lower) than the lattice temperature, the processes of phonon emission (absorption) dominate, which leads to an electric current. 

\section{Theory}

Now we develop a microscopic theory of the phonogalvanic effect. We present a brief derivation of the electron shift in 2D systems,
describe the electron states in narrow gap QWs and interaction with acoustical phonons in zinc-blende structures, calculate the electron shift and the density of electric current, and discuss the results.

\subsection{Electron shift in 2D structures}

The Bloch states of electrons with the in-plane wave vectors $\bm k$ and the spin index $s$ in a two-dimensional crystalline structure are described by the functions
\begin{equation}\label{psi_Bloch}
\psi_{\bm k s} =  U_{\bm k s} \exp(i \bm k \cdot \bm \rho) \,,
\end{equation}
where $U_{\bm k s}$ contains the Bloch amplitude and the envelope function along the quantization axis $z$ and 
$\bm \rho = (x,y)$ is the in-plane coordinate. 

To calculate the electron displacement at the quantum transition between the initial state $\nu = (\bm k, s)$ and the final state $\nu' = (\bm k', s')$ we consider that these states are described by narrow wave packets of the form
\begin{equation}\label{Psi_packets}
\Psi_{\nu} = \sum_{\bm k s} c_{\bm k s}^{(\nu)} \psi_{\bm k s}  
\end{equation}
centered at the wave vectors $\bm k$ and $\bm k'$, respectively. The displacement in the real space is determined by the difference of the position mean values in the initial and final states. For the wave packets~\eqref{Psi_packets} constructed from the Bloch functions~\eqref{psi_Bloch}, the displacement is given by 
\begin{eqnarray}
\bm R_{\nu' \nu} &=& \sum_{\bm k s} \left[ i c_{\bm k s}^{(\nu')*} \frac{d}{d \bm k} c_{\bm k s}^{(\nu')} 
+  |c_{\bm k s}^{(\nu')}|^2 \bm{\Omega}_{\bm k s}  \right] \nonumber \\
&-& \sum_{\bm k s} \left[ i c_{\bm k s}^{(\nu)*} \frac{d}{d \bm k} c_{\bm k s}^{(\nu)} 
+  |c_{\bm k s}^{(\nu)}|^2 \bm{\Omega}_{\bm k s}  \right] \,,
\end{eqnarray}
where
\begin{equation}
\bm \Omega_{\bm k s} =  i \int U_{\bm k s}^* \frac{d}{d \bm k} U_{\bm k s} d \bm r  \,,
\end{equation}
also called the Berry connection in the reciprocal space. Taking into account that the wave packets are narrow, the displacement can be expressed in terms of the gradient of the phase $\Phi_{\bm k' s, \bm k s}$ which the wave function acquires at the transition,
\begin{equation}\label{R_general}
R_{\nu' \nu} =  - \left( \nabla_{\bm k'} + \nabla_{\bm k}  \right) \Phi_{\bm k' s', \bm k s} 
+ \bm{\Omega}_{\bm k' s'} - \bm{\Omega}_{\bm k s} \,,
\end{equation}
or, equivalently, in terms of the transition matrix elements $V_{\bm k' s', \bm k s}$,
\begin{equation}
 \left( \nabla_{\bm k'} + \nabla_{\bm k}  \right) \Phi_{\bm k' s', \bm k s} = \dfrac{\operatorname{Im} \left[ V_{\bm k' s', \bm k s}^*
  \left( \nabla_{\bm k'} + \nabla_{\bm k} \right) V_{\bm k' s',\bm k s} \right]} {|V_{\bm k' s' ,\bm k s}|^2} \,. \nonumber
\end{equation}
Equation~\eqref{R_general} for the displacement of Bloch electrons at quantum transitions in crystals was obtained by Luttinger~\cite{luttinger1958} and Belinicher, Ivchenko, and Sturman~\cite{belinicher1982}.
Note that the displacement~\eqref{R_general} does not dependent on the choice of the phases of the $U_{\bm k s}$ and $U_{\bm k' s'}$ functions while the individual contributions are non-invariant.

\subsection{Electron states in narrow gap QWs}

To be specific we consider the class of $(0lh)$-oriented QWs, where $l$ and $h$ are integer numbers, which includes $(001)$-, $(011)$-, and $(013)$-grown structures. The ground electron $|e1,s \rangle$ and heavy-hole $| h1, s \rangle$ subbands at $\bm k = 0$ are described in the 6-band $\bm k$$\cdot$$\bm p$ model by the wave functions 
\begin{eqnarray} \label{psi_basis}
&&| e1,+ \rangle = f_1(z) |\Gamma_6,+1/2 \rangle + f_4(z) |\Gamma_8,+1/2 \rangle \:, \nonumber \\
&&| h1,+ \rangle = f_3(z) |\Gamma_8,+3/2 \rangle \:, \nonumber \\
&&| e1,- \rangle = f_1(z) |\Gamma_6,-1/2 \rangle + f_4(z) |\Gamma_8,-1/2 \rangle \:, \nonumber \\
&&| h1,- \rangle = f_3(z) |\Gamma_8,-3/2 \rangle \:, 
\end{eqnarray}
where $f_j(z)$ are the envelope functions, $ |\Gamma_6, m \rangle$ ($m = \pm 1/2$) and $ |\Gamma_8, m \rangle$ ($m=\pm 1/2, \pm 3/2$) are the Bloch amplitude of the $\Gamma_6$ and $\Gamma_8$ states in the center of the Brillouin zone, respectively. At 
$\bm k \neq 0$, the states are mixed. The mixing in narrow gap QWs, such as HgTe/CdHgTe, is described by the $\bm k$-linear Bernevig-Hughes-Zhang Hamiltonian~\cite{bernevig2006}
\begin{equation}\label{H_eff}
H = \left( 
\begin{array}{cccc}
\delta & {\rm i} A k_+ & 0 & 0 \\
-{\rm i} A k_- & -\delta & 0 & 0 \\
0 & 0 & \delta & - {\rm i} A k_- \\
0 & 0 & {\rm i} A k_+ & -\delta
\end{array}
\right) \,, 
\end{equation}
where $2 \delta$ is the energy gap between the $e1$ and $hh1$ subbands and $A$ is a real parameter determining the dispersion.
Admixture of excited electron and hole subbands and zero-field spin splitting are neglected in this model Hamiltonian~\cite{winkler2012,tarasenko2015}.
In QWs with symmetric confinement potential, the functions $f_1(z)$ and $f_{3}(z)$ are even while $f_4(z)$ is odd with respect to the QW center. 

Solution of the Schr\"{o}dinger equation with the Hamiltonian~\eqref{H_eff} yields the dispersion of the conduction subband
\begin{equation}
\varepsilon_{\bm k} =  \sqrt{\delta^2 + A^2 k^2} 
\end{equation}
and the wave functions
\begin{equation}\label{U_ks}
U_{\bm k s} = \frac{1}{\sqrt{2}} \left( a_k |e1 ,s \rangle -i s e^{-i s \varphi} b_k |h1 ,s \rangle \right) \,,
\end{equation}
where $s = \pm 1$,  $\varphi  = \arctan (k_y/k_x) $ is the polar angle of the wave vector $\bm k$, and 
\begin{equation}\label{akbk}
a_k = \sqrt{\dfrac{\varepsilon_k+\delta}{\varepsilon_k}} \,, \;\;
b_k = \sqrt{\dfrac{\varepsilon_k-\delta}{\varepsilon_k}} \,.
\end{equation}  

For the states in the valence subband in the electron representation one obtains
\begin{equation}
\varepsilon_{\bm k}^{(v)} = - \sqrt{\delta^2 + A^2 k^2} 
\end{equation}
and 
\begin{equation}
U^{(v)}_{\bm k s} = \frac{1}{\sqrt{2}} \left( b_k |e1 ,s \rangle +i s e^{-i s \varphi} a_k |h1 ,s \rangle \right) 
\end{equation}
with the coefficients $a_k$ and $b_k$ given by Eqs.~\eqref{akbk}.

\subsection{Electron-phonon interaction}

Consider now the deformation interaction of electrons with acoustic phonons (DA mechanism)~\cite{gantmakher1987book}. The matrix elements of the 
deformation interaction within the $\Gamma_6$ and $\Gamma_8$ bands in zinc-blend-type crystals can be presented in the form
\begin{eqnarray}\label{H_c_H_v}
\langle \Gamma_6, m | V | \Gamma_6 , m' \rangle = \Xi_c  \, {\rm Tr} \, u_{\alpha\beta} \, \delta_{mm'} \,, \nonumber \\ 
\langle \Gamma_8, m | V | \Gamma_8 , m' \rangle = \Xi_v  \, {\rm Tr} \, u_{\alpha\beta} \, \delta_{mm'} \,,
\end{eqnarray}
where $\Xi_c$ and $\Xi_v$ are the deformation-potential constants and $u_{\alpha\beta}$ are the strain tensor components. Here, we use the simplified Hamiltonian for the $\Gamma_8$ band with the single deformation-potential constant $\Xi_v$. $\Xi_v =a + (5/4)b$ in the Bir-Pikus notation~\cite{symmetrybook}, the other terms are neglected.

Importantly, in zinc-blend crystals the strain also mixes the states of the $\Gamma_6$ and $\Gamma_8$ bands~\cite{pikus1988}. Such an interband mixing in the coordinated frame $x \parallel [100]$, 
$y \parallel [0 h \bar{l}]$, $z \parallel [0lh]$, relevant for $(0lh)$-oriented QWs, is described by the matrix elements~\cite{olbrich2013}
\begin{align}\label{H_cv}
& \langle \Gamma_8, \pm 3/2 | V | \Gamma_6 , \pm 1/2 \rangle  = \sqrt{3} \, \langle \Gamma_8, \pm 1/2 | V | \Gamma_6 , \mp 1/2 \rangle \nonumber \\
& = \mp \frac{\Xi_{cv} }{\sqrt{2}} \left[ (u_{yz} \mp iu_{xz}) \cos 2\theta 
+ \left( \frac{u_{zz}-u_{yy}}{2} \pm i u_{xy} \right) \sin 2\theta \right] , \nonumber \\ 
& \langle \Gamma_8, \pm 1/2 | V | \Gamma_6 , \pm 1/2 \rangle  = \sqrt{\frac23} \left[ u_{xy} \cos 2\theta + u_{xz} \sin 2\theta \right] \,,
\end{align}
where $\theta $ is the angle between the $[001]$ and $[0lh]$ axes and $\Xi_{cv}$ is the interband deformation-potential constant~\cite{pikus1988}. Note that $\Xi_{cv}$ vanishes in centrosymmetric crystals.

The matrix elements of the scattering of electrons confined in the QW by bulk acoustic phonons can be readily calculated using the deformation Hamiltonians~(\ref{H_c_H_v},\ref{H_cv}) and the wave functions~\eqref{U_ks}. Since $|\Xi_c|, |\Xi_v| > |\Xi_{cv}|$, the scattering occurs mainly with the conservation of the spin index $s$ and by longitudinal (LA) phonons. Further, the out-of-plane component $q_z$ of the wave vector $\bm q$ of the phonons involved is typically much larger than the in-plane component $q_{\parallel} = |\bm k - \bm k'|$. In these approximations, the matrix elements of the scattering assisted by emission or absorption of LA phonons assume the form
\begin{align}\label{V_scattering}
& V_{\bm k' s, \bm k s}^{(\pm)} = \mp i \dfrac{q_z^2}{2q}  \sqrt{\dfrac{\hbar N_{q}^{(\pm)}}{2 \rho \omega_{q}}}
\Biggr[ (\Xi_c Z_{11} + \Xi_v Z_{44} ) a_{k'} a_k + \Xi_v Z_{33}   \nonumber \\ 
& \times b_{k'} b_k {\rm e}^{i s(\varphi'-\varphi)}
- i \dfrac{\sin 2 \theta}{2 \sqrt{2}}\Xi_{cv} Z_{13} (b_{k'} a_{k} {\rm e}^{i s \varphi'} - a_{k'} b_k {\rm e}^{-i s \varphi})
\Biggl] \delta_{\bm{k}',\bm{k}\mp \bm{q}_{\|}} ,
\end{align}
where $\bm q_{\parallel} = \pm (\bm k - \bm k')$, $N_{q}^{(\pm)} = N_{q} + (1\pm 1)/2$, $N_{q}$ is the phonon occupation number, $\rho$ is the crystal density, $\omega_{q} = c_s q$ is the phonon frequency, $c_s$ is the speed of longitudinal sound, and $Z_{ij}(q_z)=\int f_i(z) f_j(z) {\rm e}^{i q_z z} dz$.

\subsection{Electric current} 

The electric current caused by the shift of electrons at quantum transitions is calculated after the equation
\begin{equation}
\label{current_general}
\bm{j}=e\sum \limits_{\bm k s, \bm k' s'} W_{\bm k' s', \bm k s} R_{\bm k' s', \bm k s} \,,
\end{equation}
where
\begin{equation}
W_{\nu', \nu}  =\dfrac{2 \pi}{\hbar} \sum_{\bm{q},\pm } |V_{\nu',\nu}^{(\pm)}|^2 
f_{\nu} (1 - f_{\nu'}) \delta(\varepsilon_{\nu'} - \varepsilon_{\nu} \pm \hbar \omega_{\bm q})
\end{equation}
is scattering probability, $R_{\nu'\nu}$ is the shift given by~\eqref{R_general}, and $f_{\nu}$ is the electron distribution function.

We assume that electron thermalization governed by electron-electron collisions is established at a time scale which is much shorter than the energy relaxation time determined by inelastic electron-phonon scattering. As a result, the electron distribution is quasi-equilibrium and can be described by the Fermi-Dirac function $f_{\nu} = \{\exp[(\varepsilon_{\nu} - \mu)/k_B T_e] + 1\}^{-1}$, where $T_e$ is the effective electron temperature and $\mu$ is the chemical potential. The phonon distribution is described by the Bose-Einstein function $N_{\bm q} = [\exp(\hbar \omega_{\bm q}/k_B T) - 1]^{-1}$ with the lattice temperature $T$. 
At $\Delta T = T_e - T \ll T$, the difference between the rates of photon emission and absorption is given by
\begin{equation}
\left[f_{\nu'} (1 - f_{\nu}) N_q^+-f_{\nu}(1-f_{\nu'}) N_q^- \right]_{\varepsilon_{\nu'} =
\varepsilon_\nu+\hbar \omega_q} \approx f_{\nu'} (1-f_{\nu}) \dfrac{\Delta T}{T} \,.
\end{equation}

To proceed further, we calculate the current contributions $\bm j^{(1)}$ and $\bm j^{(2)}$ related to the phase gradients and the Berry connections in Eq.~\eqref{R_general}, respectively, separately. However, we stress that only the sum of them, 
$\bm j = \bm j^{(1)} + \bm j^{(2)}$, is invariant with respect to the wave function choice and is a physical observable.

The contribution to the current stemming from the phase gradients is determined by the vectors
\begin{equation}
\label{shift_phase_general}
\bm{\Lambda}^{(\pm)}  = \left\langle V_{\bm k' s', \bm k s}^{(\pm)*}
  ( \nabla_{\bm k'} + \nabla_{\bm k} ) V^{(\pm)}_{\bm k' s',\bm k s} \right\rangle ,
\end{equation}
where the angle brackets denote averaging over the directions of the wave vectors $\bm{k}$ and  $\bm{k}'$.
For the matrix elements~\eqref{V_scattering}, they have the form
\begin{multline}
\label{shift_phase}
\bm{\Lambda}^{(\pm)} = 
\frac{\hat{\bm x} \,  \sin 2\theta }{ 32 \sqrt{2}}
\frac{ \hbar q_z^2 Z_{13} (q_z) \Xi_{cv}  N_{q}^{(\pm)} }{ \rho \omega_{q}}
\Biggr\{
[\Xi_c Z_{11}(q_z)+\Xi_{v} Z_{44}(q_z)]\times\\
\biggl[
a_{k'}a_{k}a_{k'}\left(\dfrac{d}{d k}+\dfrac{1}{k}\right)b_{k}
- a_{k'}a_{k}a_{k}\left(\dfrac{d}{d k'}+\dfrac{1}{k'}\right)b_{k'}
- a_{k'} b_{k} a_{k'} \dfrac{d a_k}{d k} 
+ a_{k} b_{k'} a_{k} \dfrac{d a_{k'}}{d k'} 
\biggr] 
\\
- \Xi_{v} Z_{33}(q_z)
\biggl[
b_{k'} b_{k} b_{k'}\dfrac{d a_{k}}{d k}
- b_{k'} b_{k} b_{k}\dfrac{d a_{k'}}{d k'}
- b_{k'} b_{k'} a_{k}\left(\dfrac{d}{d k}+\dfrac{1}{k}\right)b_{k}
+ b_{k} b_{k} a_{k'}\left(\dfrac{d}{d k'}+\dfrac{1}{k'}\right)b_{k'}
\biggr]
\Biggl\} ,
\end{multline}
where $\hat{\bm x}$ is the unit vector along the $x$ axis. This results in the following contribution to the electric current
\begin{equation}
\label{phase_current}
j^{(1)}_x= \frac{e \hbar \sin 2\theta}{64 \sqrt{2} \pi^2 } \frac{\Xi_{cv} \Delta T}{\rho A^3 d^3}
\int\limits_{\delta}^{+\infty} \left[
\left(\Xi_c \zeta_{1}+\Xi_v \zeta_{4} \right) a_k^4+
\Xi_v \zeta_{3} \left(\frac{a_k^4+2 a_k^2 b_k^2-b_k^4}{2}\right) \right] \dfrac{f_k(1-f_k)}{T} d \varepsilon_k \,,
\end{equation}
where $d$ is the QW width, $\zeta_{j}$ are dimensionless parameters determined by the functions of size quantization,
\begin{equation}
\zeta_{j} = d^3 \int\limits_{-\infty}^{+\infty} dq_z q_z^2 Z_{13}(q_z) Z_{jj}(q_z) = 2 \pi d^3 \int\limits_{-\infty}^{+\infty} \frac{d \, f_1(z) f_3 (z)}{dz} \frac{d \, f_j^2(z)}{d z} dz \:,
\end{equation}
and it is assumed that the energy of phonons involved is much smaller than the mean electron energy.
\begin{figure}[H]
\begin{center}
\includegraphics[width=0.5\paperwidth]{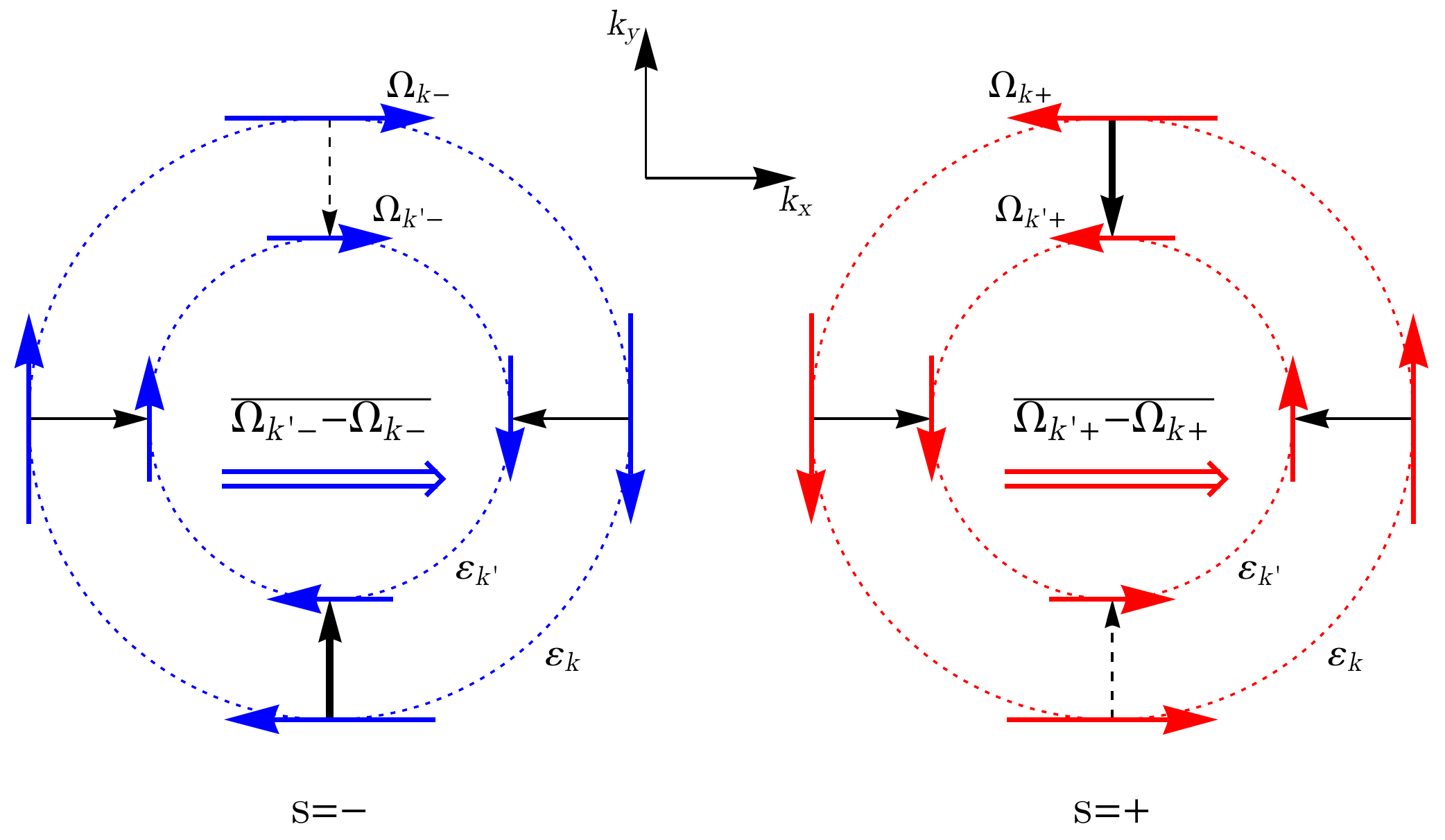}
\caption{
\label{omega_pic}
Distribution of the Berry connection $\bm{\Omega}_{\bm{k}s}$ in the $\bm k$ space in the spin subbands $s = "\pm"$ plotted after Eq.~\eqref{BC_in_k}. Red and blue dotted circles show isoenergetic curves. Black arrows sketch the electron transitions with the emission of phonons. Asymmetry of the electron transitions in the spin and momentum spaces (shown by black arrows of different thickness) at energy relaxation leads to a variation of the mean value of $\bm{\Omega}_{\bm{k}s}$, i.e., to the contribution $\bm {j}^{(2)}$ to the shift current.
}
\end{center}
\end{figure}

The second contribution to the current is determined by the Berry connection which, for the wave functions~\eqref{U_ks}, 
has the form
\begin{equation}\label{BC_in_k}
\bm \Omega_{\bm k s} = \frac{s b_k^2}{2 k} \, \hat{\bm{z}} \times \hat{\bm{k} } \,,
\end{equation}
where $\hat{\bm{z}}$ and $\hat{\bm{k}}$ are the unit vectors along $\bm z$ and $\bm k$, respectively. The Berry connection lies in the QW plane and points in the direction perpendicular to $\bm k$. Its direction depends on the spin index $s$ and its amplitude is proportional to $b_{k}^2/k$. The distribution of the Berry connection in the $\bm k$ space for the both spin subbands is shown in Fig.~\ref{omega_pic}. 
Black arrows in the same figure sketch the electron transitions with the initial $\bm k$ and final $\bm k'$ wave vectors with the emission of phonons. Owing to the term $\propto \Xi_{cv}$, the rate of such transitions is asymmetric in the $\bm k$ space, 
which is illustrated by the arrows of different thickness. Importantly, the sign of the asymmetry also depends on the spin index 
$s$. Therefore, the value of $\bm{\Omega}_{\bm{k}'s}-\bm{\Omega}_{\bm{k}s}$ summed up over the initial and final wave vectors is the same for the spin subbands and contributes to the shift current.
Straightforward calculations yields the Berry connection contribution to the current  
\begin{equation}
\label{omega_current}
j_x^{(2)} = \frac{e \hbar \sin 2\theta}{64 \sqrt{2} \pi^2 } \frac{\Xi_{cv} \Delta T}{\rho A^3 d^3}
\int\limits_{\delta}^{+\infty}  \left[ (\Xi_c \zeta_{1}+\Xi_v \zeta_{4}) \left(\dfrac{a_k^4-2a_k^2 b_k^2-b_k^4}{2}\right)-
\Xi_v \zeta_{3} b_k^4
\right] \dfrac{f_k(1-f_k)}{T} d \varepsilon_k \:.
\end{equation} 

Finally, summing up the partial contributions~\eqref{phase_current} and~\eqref{omega_current} we obtain the shift electric 
current caused by the energy relaxation of electrons in the conduction subband
\begin{equation}
\label{total_current}
j_x = \frac{e \hbar \sin 2\theta}{32 \sqrt{2} \pi^2 } \frac{\Xi_{cv} \Delta T}{\rho A^3 d^3}
\int\limits_{\delta}^{+\infty}  \left[
\left(\Xi_c \zeta_{1}+\Xi_v \zeta_{4} \right) \left(2 +\frac{\delta}{\varepsilon_k}\right)+\Xi_v \zeta_{3} \left(2 -\frac{\delta}{\varepsilon_k}\right) \right] \frac{\delta}{\varepsilon_k} \dfrac{f_k(1-f_k)}{T} d \varepsilon_k  \:.
\end{equation}

The shift current emerges also in $p$-type QWs at energy relaxation of holes. Similar calculations carried out for the case when the chemical potential lies in the valence subband give also Eq.~\eqref{total_current} where the integral is taken from $-\infty$ to $-\delta$.
Equation~\eqref{total_current} is the main result of this paper. Below, we discuss it and analyze for some particular cases.

\section{Discussion}

The shift current depends on the band structure parameters, the QW crystallographic orientation, and the carrier distribution function.
Its dependence on the QW orientation is given by $\sin 2 \theta$. The case of $\theta = \pi n/2$ with integer $n$, when the current vanishes, corresponds to $\{001\}$-grown QWs. Such structures have high symmetry described by the $D_{2d}$ or $C_{2v}$ point groups depending on whether the confinement potential is symmetric or asymmetric. The structures contain two mirror planes orthogonal to each other and to the QW plane, which forbids the generation of an in-plane electric current by a scalar perturbation of the system, such as breaking the thermal equilibrium between electrons and phonons. At $\theta \neq \pi n/2$, the QW structures have lower symmetry (our model Hamiltonian corresponds to the $C_s$ point group, the actual symmetry can be even lower). They belong to the class of pyroelectric structures with the in-plane component of the polar vector, and the symmetry allows the current  generation.

The integral in Eqs.~\eqref{total_current} can be calculated analytically in the case of low temperature and degenerated electron gas when ${f_k(1-f_k)/T\approx \delta(\varepsilon_k - \varepsilon_F)}$, where $\varepsilon_F$ is the Fermi level measured from the center of the band gap. Then, the expression for the current assumes the form
 \begin{equation}\label{current_degenerate}
j_x = \frac{e \hbar \sin 2\theta}{32 \sqrt{2} \pi^2 } \frac{\Xi_{cv} \Delta T}{\rho A^3 d^3} \frac{\delta}{\varepsilon_F^2}
\left[ (\Xi_c \zeta_{1}+\Xi_v \zeta_{4} ) (2 \varepsilon_F + \delta)+\Xi_v \zeta_{3} (2 \varepsilon_F - \delta) \right] \Theta(|\varepsilon_F| - \delta) \,,
\end{equation}
where $\Theta$ is the Heaviside step function. Equation~\eqref{current_degenerate} is valid for both $n$-type and $p$-type structures.
 
A prominent example of narrow gap QWs with the zinc-blend lattice is HgTe/CdHgTe structures.
Therefore, we make quantitative calculations of the shift current for the parameters relevant to HgTe~\cite{adachi2004}:
the crystal density $\rho=8$~g/cm$^3$, the deformation-potential constants $\Xi_c=-3.8$~eV and $\Xi_v=-2$~eV,
and the ratio $\Xi_{cv}/\Xi_c = 1/3$ (this ratio is not known for HgTe, we use the ratio for GaAs~\cite{pikus1988}).
The band gap $\delta$ and overlap integrals $\zeta_j$ are calculated numerically in the $\bm{k}\cdot \bm{p}$-model~\cite{dantscher2015}.

\begin{figure}[H]
\begin{center}
\includegraphics[width=0.5\paperwidth]{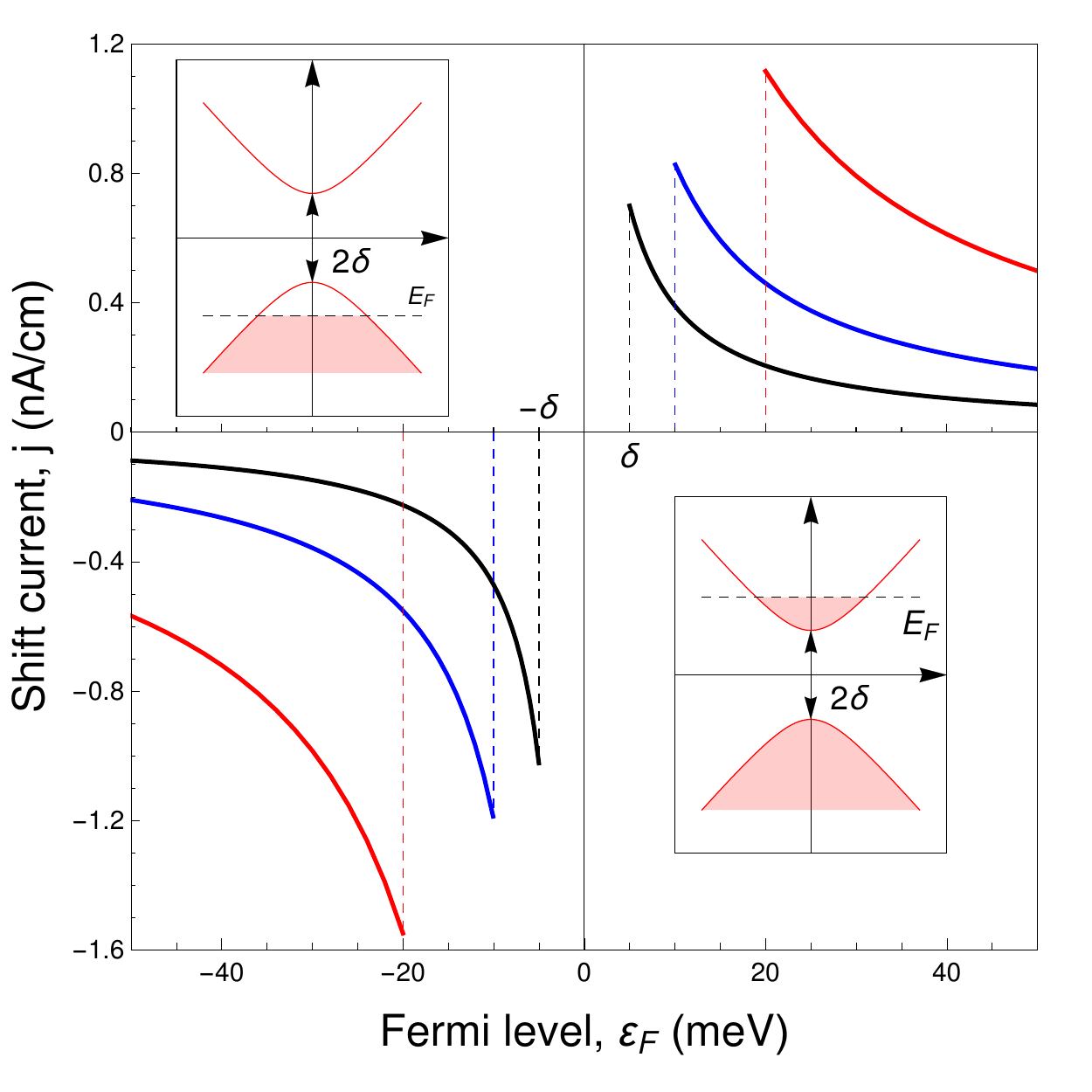}
\caption{
\label{HgTe}
Dependence of the shift current on the Fermi level position calculated after Eq.~\eqref{current_degenerate} for different
band gaps $\delta$, the parameters relevant to the HgTe/CdHgTe QWs, $\Delta T = 10$~K, and $\theta=\pi/4$. Black line correspond to $\delta=5$~meV (QW width $5$~nm), blue line to $\delta=10$~meV (width $5.6$~nm) and red line to $\delta=20$~meV (width $6.0$~nm).
}
\end{center}
\end{figure}

Figure~\ref{HgTe} shows the dependence of the shift current on the Fermi level position calculated after Eq.~\eqref{current_degenerate}. Curves of different colors correspond to the QWs with different band gap which can be controlled in HgTe/CdHgTe structures by the QW thickness.
Moreover, in HgTe/CdHgTe structures the Fermi level position can be tuned from the conduction 
to the valence subband by applying the gate voltage. Figure~\ref{HgTe} shows that the current flows in the 
opposite directions in $n$-type and $p$-type structures. 
At small densities of electrons or holes, when the 
Fermi level $\varepsilon_F$ is close to $\pm \delta$, respectively,  the electric currents reach the values 
 \begin{equation}
j_{c,v} = \pm \frac{e \hbar \sin 2\theta}{32 \sqrt{2} \pi^2 } \frac{\Xi_{cv} \Delta T}{\rho A^3 d^3}  
\left[ (2 \pm 1) (\Xi_c \zeta_{1}+\Xi_v \zeta_{4} ) + (2 \mp 1) \Xi_v \zeta_{3} \right] .
\end{equation}
%

%

The theory of the shift current is developed here for a small deviation between the effective electron and lattice temperatures.
It gives $j \sim $~nA/cm for $\Delta T = 10$~K and the energy relaxation by acoustic phonons. If fact, the current magnitude is proportional to the rate of energy transfer between the electron and lattice subsystems. The estimations above correspond to the 
energy transfer rate $\dot{E} \sim 0.3$~mW/cm$^2$. In real experiments on electron gas heating by THz or optical pulses, $\dot{E}$ can be orders of magnitude larger giving rise to much stronger electric current.

\section{Summary}

We have shown that the energy relaxation of a heated (or cooled) electron gas in low-symmetry quantum wells by phonons drives a direct electric current and developed a microscopic theory of such a phonogalvanic effect. 
The current originates from the shifts in the real space that the Bloch electrons experience in the course of quantum transitions at emission or absorption of phonons. When the thermal equilibrium between the electron and phonon subsystems is broken and the subsystems are characterized by different effective temperatures, the shifts get a preferable direction, which results in a net electric current. We have calculated the electric current for the deformation mechanisms of electron-phonon interaction in narrow band gap zinc-blende quantum wells.

This work was supported by the Government of the Russian Federation (contract \# 14.W03.31.0011 at Ioffe Institute).
G.V.B also acknowledges the support from the BASIS foundation.
    
\bibliography{scref}
  
\end{document}